\documentclass[printer]{aa}
\usepackage{natbib}
\usepackage{graphicx}
\usepackage[caption=false]{subfig}
\usepackage[english]{babel}
\usepackage{lscape}
\usepackage{amssymb}
\usepackage{amsmath}
\usepackage{auto-pst-pdf}
\graphicspath{{./../figures_article/}}
\newcommand{\gppr}{\stackrel{>}{\scriptstyle \sim}}
\newcommand{\gappr}{\raisebox{-0.4ex}{$\gppr$}}
\newcommand{\lppr}{\stackrel{<}{\scriptstyle \sim}}
\newcommand{\lappr}{\raisebox{-0.4ex}{$\lppr$}}

\newcommand{\Mwd}{\mbox{$M_\mathrm{wd}$}}
\newcommand{\Msec}{\mbox{$M_\mathrm{sec}$}}
\newcommand{\Mone}{\mbox{$M_\mathrm{1}$}}
\newcommand{\Mtwo}{\mbox{$M_\mathrm{2}$}}

\newcommand{\Msun}{\mbox{$\mathrm{M}_{\odot}$}}
\newcommand{\Rsun}{\mbox{$R_{\odot}$}}

\begin{document}

\title{White dwarf masses in cataclysmic variables}
\titlerunning{WD masses in CVs}
\author{T.P.G. Wijnen\inst{1,2}, M. Zorotovic\inst{1}, M.R. Schreiber\inst{1,3}
}
\authorrunning{T.P.G. Wijnen et al.}
\institute{Instituto de F\'isica y Astronom\'ia, Facultad de Ciencias, Universidad de Valpara\'iso, Valpara\'iso, Chile 
\and Department of Astrophysics/IMAPP, Radboud University Nijmegen, P.O. Box 9010, 6500 GL Nijmegen, The Netherlands \\
\email{thomas.wijnen@astro.ru.nl}
\and{ICM nucleus on protoplanetary disks, Universidad de Valpara\'iso, Av. Gran Breta\~na 1111, Valpara\'iso, Chile}
\offprints{T.P.G. Wijnen}
}
\date{Received 12 November 2013/ Accepted 8 March 2015}

\abstract{The white dwarf (WD) mass distribution of cataclysmic variables (CVs) has recently been found to dramatically 
disagree with the predictions of the standard CV formation model. The high mean WD mass among CVs is not imprinted in the  
currently observed sample of CV progenitors and cannot be attributed to selection effects. Two possibilities have been put 
forward to solve this issue: either the WD grows in mass during CV evolution, or in a significant fraction of cases, CV 
formation is preceded by a (short) phase of thermal time-scale mass transfer (TTMT) in which the WD gains a sufficient amount of mass.}
{Here we investigate if and under which conditions a phase of TTMT before CV formation or mass growth in CVs can bring theoretical predictions and observations into agreement.}
{We employed binary population synthesis models using the \emph{binary\_c/nucsyn} code
  to simulate the present intrinsic CV population. To that end we incorporated
  aspects specific to CV evolution such as an appropriate
  mass-radius relation of the donor star and a more detailed prescription for the critical mass
  ratio for dynamically unstable mass transfer. We have also implemented a
  previously suggested wind from the surface of the WD during TTMT and tested
  the idea of WD mass growth during the CV phase by arbitrarily 
changing the accretion efficiency. We compare the model predictions of the TTMT and the 
mass growth model with the characteristics of CVs derived from observed samples.}
{We find that mass growth of the WDs in CVs fails to reproduce the observed WD
  mass distribution. In the case of TTMT, we are able to produce a large number
  of massive WDs if we assume significant mass loss from the surface of the WD
  during the TTMT phase. However, the model still produces too many CVs with 
  helium WDs. Moreover, the donor stars are evolved in many of these post-TTMT CVs, which contradicts the observations.}
{We conclude that in our current framework of CV evolution neither TTMT nor WD mass growth 
can fully explain either the observed WD mass or the period distribution in CVs.}
\keywords{accretion, accretion discs -- instabilities -- stars: novae, cataclysmic variables -- stars: binaries: close}

\maketitle

\section{Introduction}

The class of compact binary stars comprises a great diversity of stellar
objects and phenomena in the galactic zoo. They are very important probes of
our comprehension of stellar evolution in general and mass 
transfer in particular. Cataclysmic variables (CVs) are compact binaries
consisting of a white dwarf (WD) and a low-mass main sequence (MS) star that
transfers mass to the WD by Roche-lobe overflow. CVs have been
investigated for several decades, but their formation and evolution is still
not fully understood. It is generally accepted that CVs result from wide
binaries evolving into a common envelope (CE) structure, from which the core
of the giant, that is, the primary, remains as a WD and in which the separation
decreases significantly by means of drag forces within the envelope
\citep{paczynski76-1}. After the envelope is expelled, the orbit of the
detached post-common-envelope binary (PCEB) is further reduced through the
loss of orbital angular momentum by gravitational radiation (GR) and magnetic
breaking (MB). When the orbit is sufficiently close, the accompanying MS star
fills its Roche lobe, and if the resulting mass transfer is stable, a CV is
born.  

According to this CV formation scenario, WDs in the newly formed CVs should
have a mass distribution that is similar to the mass distribution of single WDs, if
not shifted towards lower masses by an early expulsion of the envelope,
which prematurely terminates the mass growth of the giant's core. This naive
expectation of on average low WD masses has been confirmed by binary
population models of CVs, \citet{politano96-1} for instance predicted a mean WD mass
of $0.49\,\Msun$ for the primaries of CVs. However, measurements of WD masses
in CVs have been in the range of [0.8-1.2]\,\Msun\,(e.g., \citealt{warner73-1,
warner76-1, ritter76-1, robinson76-1}), which is significantly higher than
predicted. This discrepancy between the observed and expected mean WD mass in
CVs has been successfully interpreted as a selection effect by
\citet{ritter+burkert86-1}. Simply speaking, the idea is that  the higher the
WD mass, the more energy is released per accreted unit mass and the more
extended is the accretion disk around the WD. Thus, CVs with massive WDs are
(on average) significantly brighter and much easier to be discovered. However, \citet{zorotovicetal11-1} recently showed that this previous  explanation does no
longer hold. They showed that the observed WD mass distribution of 
faint CVs (dominated by the emission from the WD instead of by
that from the accretion disk) should
be biased towards {\em{low-mass}} WDs, while, as shown by
\citet{littlefairetal08-1} and \citet{savouryetal11-1}, the measured mean WD mass for these systems still
remains at $\sim0.8\,\Msun$. 

Thus the standard model of CV evolution might miss an important
ingredient. \citet{zorotovicetal11-1} suggested two possibilities. First, WDs in
CVs may gain mass through accretion of transferred matter over a nova cycle if
less mass is expelled during the eruptions than is accreted between them.
This contradicts standard theories for nova outbursts \citep{prialnik86-1,
prialnik+kovetz95-1,yaronetal05-1}. 
WD masses can only grow in CVs if no mixing of core matter 
with accreted matter is assumed 
\citep{williams13-1, starrfield15-1}, which is unrealistic.
However, given the large discrepancy
between the observed and predicted WD mass distributions, we need to
investigate all possible scenarios. Therefore, we here present binary
population models that include different mass accretion efficiencies in CVs.  

The second possible explanation put forward by \citet{zorotovicetal11-1} is
that a large number of CVs descend from binaries with initially more massive
secondary stars. This implies a preliminary phase of thermal time-scale mass
transfer (TTMT) in which the mass of the WD grows through stable hydrogen
burning on its surface \citep{schenkeretal02-1}. At that stage, the system
might be observed as a super-soft X-ray source \citep{kahabka+vandenheuvel97-1}. 
A small sample of UV observations has shown that 10-15\,\% of CVs accrete CNO processed
material, indicating that the companion has been stripped of its external
layers by a previous phase of TTMT \citep{schenker+king02-1,gaensickeetal03-1}. 
These companions therefore appear to be more evolved than a single MS star of the 
same mass. While binary population models of new-born CVs with evolved donor stars  
have been applied in the past \citep[e.g.][]{dekool92-1,
baraffe+kolb00-1,podsiadlowskietal03-2, kolb+willems05-1}, a systematic
study of the impact of TTMT on the WD mass distribution of CVs is
missing. Here we fill this gap by using updated binary population models and
investigate whether a large number of CVs descending from a
phase of TTMT might explain the large masses of CV primaries.   

\section{Code}
\label{sec:code}

We simulated the evolution of a large number of binaries and selected those
systems that evolve into a CV within the age of the Galaxy. Then we
quantitatively investigated their characteristics and the evolutionary channel
by which they have formed. We used the population nucleosynthesis code
\emph{binary\_c/nucsyn} of \citet{izzardetal04-1, izzardetal06-1,
  izzardetal09-1} based on the binary star evolution code of
\citet{hurleyetal02-1}.  

To simulate CVs, some specific aspects of their evolution have 
to be taken into account. In particular, the implementation and treatment 
of the stability of mass transfer, the mass-radius relation of the donor star, 
and the fate of the transferred matter have to be considered carefully.  
We address these issues in detail below. 

\subsection{Mass transfer}
\label{sec:stab_mt}
The stability of mass transfer is determined by the change of the radius with respect
to the Roche lobe. If the adiabatic response of the donor is unable to retain the star
within its Roche lobe, mass transfer will occur in a dynamically unstable way and will
probably lead to a CE. However, if the star is able to restore hydrostatic equilibrium,
mass transfer is determined by the thermal readjustments of the star. If the new thermal 
equilibrium radius exceeds the Roche lobe radius, mass transfer is driven by readjustments 
of the star on the thermal time-scale. Otherwise, the binary is stable against mass transfer
and mass transfer is driven by angular momentum loss or nuclear evolution. 

\begin{figure}[!t]
\begin{center}
    \includegraphics[width=0.49\textwidth]{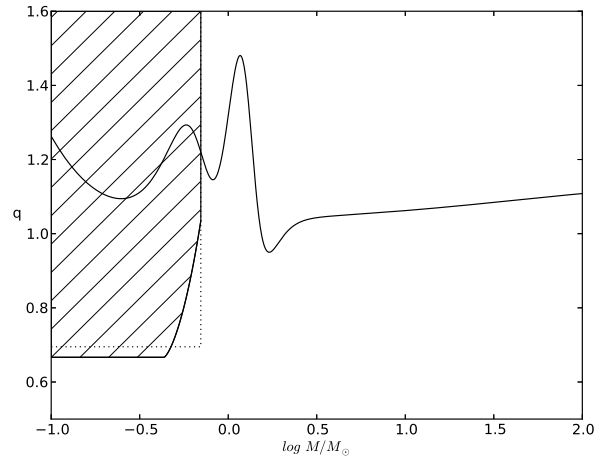}
    \caption{Different regimes of stability against mass transfer for MS
      donor stars, depending on the mass ratio and the mass of the donor. The
      hatched region corresponds to the critical mass ratios for which, according to
      \citet{politano96-1}, the mass transfer becomes dynamically
unstable as a function of the donor's mass. The dotted
      line indicates the shape of this regions if one uses the condition that
      is originally in the code (see text). The solid curve represents the
      critical mass ratio for zero-age MS stars of solar composition above
      which mass transfer occurs on the thermal time-scale. This mass ratio is
      derived from the thermal mass-radius exponent by using the zero-age MS
      mass-radius relation implemented in the code \citep{toutetal96-1} and
      assuming that mass transfer is conservative.\label{fig:webbink}}
\end{center}
\end{figure}

The adiabatic mass-radius exponent for low-mass MS donors is very sensitive to the depth of
their convective envelope. For low-mass MS stars ($\lesssim 0.7\,\Msun$) the envelope is deeply convective
and the donor star is no longer able to restore hydrostatic equilibrium in response to mass loss, 
except for very low mass ratios. Therefore, the adiabatic mass-radius exponent and corresponding
critical mass ratio decrease steeply around $0.7\,\Msun$. Figure\,\ref{fig:webbink} shows an analytic
fit from \citet{politano96-1} to the critical mass ratio for dynamically unstable mass transfer
(hatched region): 
\begin{equation}\label{eq:qad}
q_{\rm cr}= \begin{cases}
\frac{2}{3} &\text{if $\Mtwo \le 0.4342$}\\
2.244(\Mtwo-0.4342)^{1.364} + \frac{2}{3} &\text{if $0.4342 \le \Mtwo \le 0.7$}
\end{cases}
,\end{equation}
where $\Mtwo$ represents the mass of the donor star in solar masses and the mass ratio is 
defined as $q \equiv \frac{\Mtwo}{\Mwd}$. This fit is based on the adiabatic mass-radius 
exponent as determined by detailed model calculations of \citet{hjellming89-1} and is valid 
for conservative mass transfer. We used this more accurate prescription instead of 
the standard prescription in \emph{binary\_c/nucsyn}, that is, a constant $q_{\rm cr}$
of $0.695$ for $\Mtwo \lesssim 0.7\,\Msun$ (which is also shown in Fig.\,\ref{fig:webbink} 
for comparison). For masses above $0.7\,\Msun$, mass transfer is assumed to be dynamically 
stable if $q < 1.6$ in our simulations \citep{deminketal07-1}. This line is not shown in Fig.\,\ref{fig:webbink}.

Since \emph{binary\_c/nucsyn} does not follow the thermal mass-radius exponent
of the donor (\citealt{hurleyetal02-1}, section 2.6.3) and mass transfer might
not be conservative during TTMT, we used the mass-transfer rate to distinguish
between the CV phase and a phase of TTMT. We identify a binary as a system
with TTMT if the primary is a WD, consisting of either helium (He),
carbon-oxygen (C/O), or oxygen-neon (O/Ne), the donor is a MS star and the
mass-transfer rate is higher than the limit for stable hydrogen burning, as
described in \citet{mengetal09-1}. Furthermore, the mass of the WD has to
increase by at least 0.01\,$\Msun$ during the TTMT phase, otherwise we do not
define the emerging CV as a post TTMT system. Likewise, if the mass-transfer
rate is below the limit for stable hydrogen burning, the primary is a WD, 
and the donor is a MS star, we identify the system as a CV. This definition of 
CVs may include some long orbital period systems with massive secondaries
($\gappr\,0.7\,\Msun$) and lower mass WDs that just started thermally unstable 
mass transfer but did not yet reach the limit for stable hydrogen burning. We count these systems as CVs keeping in mind that they are apparently 
different from typically observed CVs.

\subsection{Mass-radius relation for CV donors}

Mass transfer can force the donor star out of thermal equilibrium when the
time-scale of mass transfer is shorter than the thermal time-scale of the
donor. In the case of low-mass donors, the mass transfer is driven by the loss
of angular momentum due to MB and/or GR. We subtracted the angular momentum loss
due to MB directly from the orbit, assuming the orbit and spin are
coupled. 
We used the prescription of \citet{hurleyetal02-1} for MB. MB is assumed to be active until the donor star becomes fully convective
\citep{rappaportetal83-1}. For single zero-age MS (ZAMS) stars and in detached binaries, this occurs when the mass 
is $\sim  0.35\,\Msun$ and MB is probably 
reduced significantly in an abrupt manner \citep{schreiberetal10-1}. 
This disrupted MB scenario can explain the 
observed gap between $\sim$\,2 and 3 hours in the orbital period distribution of CVs
\citep[e.g.][]{spruit+ritter83-1}. In semi-detached 
binaries that contain a secondary with a radiative core, 
the strong angular momentum loss and the resulting high mass transfer rates
due to MB drive the MS donor out of thermal equilibrium, which causes the
radius of the donor star to exceed its thermal equilibrium radius. 
As a result of this bloating and readjustments of the donor on relatively long thermal
time-scales, the stellar structure of the donor corresponds to the stellar
structure of a more massive single star on the MS \citep{howelletal01-1,
  knigge06-1}. In other words, the mass of the donor is lower than that of
a MS star with the same radius and stellar structure. Therefore 
the mass-transferring donor in a CV becomes fully convective 
at a lower mass than its MS counterparts 
in detached binaries or single stars. As a result, the dynamo
mechanism responsible for MB remains active for donor masses $\gtrsim
[0.2-0.26]\,\Msun$  \citep{mcdermott+taam89-1, howelletal01-1,
  pattersonetal05-1}. 
When MB becomes inefficient at an orbital period of $\sim3$\,hr, 
the donor has time to relax its radius to its
equilibrium value, which is lower than the value of the Roche-lobe radius. The binary, having become a detached system, evolves towards shorter orbital periods driven by GR alone. At an orbital period of $\sim2$\,hr, 
mass transfer starts again, but at a much lower rate.
In other words, the standard scenario for CV evolution explains the deficit 
of CVs in the period gap by predicting that CVs pass the gap as detached
systems.   

Because of this standard theory of CV evolution, the mass-radius
relation of the donor star is of crucial importance for properly simulating the standard model of CV evolution, as the disrupted MB scenario only works
if secondaries above the gap have a larger radius than their MS radius. Not
assuming an increased radius above the gap would not allow simulating the
orbital period gap seen in the observed distribution of CVs. Consequently, we
would not be able to separate systems below and above the gap, which, as we
show below, might be imperative for understanding the WD mass distribution in
CVs. To account for the larger radius when the donor is out of
thermal equilibrium, we implemented the mass-radius relation for low-mass MS
donors in CVs, as deduced by \citet{kniggeetal11-1}. To establish a smooth
transition between the equilibrium radius ($R_{\rm2,eq}$) given by
\emph{binary\_c/nucsyn} and the increased radius for CV donor stars
($R_{\rm2,CV}$), we define the factor by which the radius is increased 
with respect to its thermal equilibrium value, as a function of the current mass of the donor star $m$,  
\begin{equation}\label{eq:factor}
f(m) = \frac{R_{\rm 2,CV}(m)}{R_{\rm 2,eq}(m)}
,\end{equation}
and let the radius grow exponentially with time towards the fully inflated value 
(i.e. $R_{\rm 2,CV}$) given by \citet{kniggeetal11-1}. The equilibrium radius excess
($f_{\rm exc}$) and current radius ($R_{\rm 2}$) as a function of time and mass are thus given by

\begin{equation}\label{eq:inflation1}
f_{\rm exc}(t,m) = f(m) + (1-f(m))e^{-\frac{t}{\tau}},
\end{equation}

\begin{equation}\label{eq:inflation2} 
 R_{\rm 2}(t,m) = f_{\rm exc}(t,m) R_{\rm 2,eq}(m),
\end{equation}

\noindent where $t$ is the time since the donor filled its Roche lobe and $\tau$ is the time scale for 
angular momentum loss in CVs, typically $10^{7}$ years \citep{davisetal08-1}. In accordance 
with the value for $M_{\rm conv}$ from
\citet{kniggeetal11-1}, we assume MB is disrupted at $0.20\,\Msun$ for CVs.  
When MB becomes inefficient, the donor has time to relax its radius to its
equilibrium value. 
In this case, we use Eqs. (\ref{eq:factor}) and (\ref{eq:inflation2}), replace $R_{\rm 2,CV}$ with the
radius the star had just before it detached and $R_{\rm 2,eq}$ with the radius
as described in \citet{kniggeetal11-1} for donors below the period gap. 
This radius decrease is implemented in the code in a similar fashion as the increase 
(Eq. \ref{eq:inflation1}), that is, decreasing the radius exponentially, where now
\begin{equation}\label{eq:factordec}
 f_{\rm exc}(t,m) = 1 - (1-f(m))e^{-\frac{t}{\tau}}. 
 \end{equation}
We assume that donor stars whose initial mass is lower than $0.35\,\Msun$ do not
experience efficient MB, analogous to single ZAMS stars. Therefore, we did not
inflate the radius for these donors, but only used the radius for CV donors
below the gap. As pointed out by \citet{kniggeetal11-1}, the power-law
approximation for the mass-radius relation of MS donors breaks down for masses
$\le 0.05 \,\Msun$. We therefore only considered CVs with a secondary more
massive than $0.05\,\Msun$.

\subsection{CVs with evolved donor stars}
\label{sec:evolved_donors}
While the assumed mass radius relation of CVs is reasonable for CVs with MS
donor stars, we have to take into account that CVs from binaries with
more massive secondaries, such as those that passed through a phase of TTMT, 
may have significantly evolved donor stars. 

As has been shown by \citet{podsiadlowskietal03-2}, the orbital period
evolution of CVs with donor stars that have a central hydrogen fraction 
below 0.4 significantly differs from those with less evolved secondary stars. 
In particular, CVs with evolved donors do not become fully convective at an
orbital period of $\sim3$\,hr as ``normal'' CVs do, but at possibly much
shorter periods \citep{baraffe+kolb00-1,podsiadlowskietal03-2}. 
In fact, they may even bounce back and evolve to longer periods before 
becoming fully convective.
As a consequence, CVs with evolved donors pass the period gap as
accreting systems with a bloated donor star that is significantly hotter than
a typical CV donor at the same period. Such systems have indeed 
been found
\citep{thorstensenetal02-1,littlefairetal06-1,thorstensen13-1,rebassa-mansergasetal14-1}.
Thus, if many of them are produced in our simulations, then the predicted orbital 
period distribution could be significantly affected. 
Incorporating evolutionary tracks for CVs with evolved secondaries in 
our simulations is beyond the scope of this paper. Instead, we simply keep 
track of the central hydrogen fraction of the CV donors, derived from their 
initial mass and current age, and determine the fraction of systems in our simulation 
that contain evolved donors as defined by \citet{podsiadlowskietal03-2}, that is, with donor 
stars that have a central hydrogen fraction below 0.4. 

To evaluate whether a given predicted fraction of evolved systems would disagree 
with the observations, we derived an upper limit from the observed period distribution. 
Assuming that all observed CVs in the gap are CVs with evolved donors (this is
a strict upper limit as some CVs with un-evolved donors are born in the gap) 
and taking into account that the number of CVs at the lower edge of the gap in the 
observed distribution increases roughly by a factor 
of 3-5 \citep{gaensickeetal09-1,kniggeetal11-1}, we estimate an upper
limit for the fraction of CVs with evolved donors 
(that evolve through the gap as accreting systems) of $\sim20-30$ 
per cent. If this fraction is exceeded, one would expect more CVs in the gap
and a smaller increase of the number of CVs at the lower edge of the gap 
than is observed.
   
\subsection{Modelling the mass-transfer rate}

\emph{Binary\_c/nucsyn} is a parametrized stellar evolution code and does not 
calculate the various mass-radius exponents that are discussed in 
Sect. \ref{sec:stab_mt}. Instead the calculation of the rate at which mass is
transferred to the accretor, $\dot{M}_{\rm tr}$, contains a numerical factor
to ensure that the mass-transfer rate is (numerically) steady (see Eq. (59) from
\citealt{hurleyetal02-1}). However, this factor is too small to let the star
follow its Roche lobe within a few per cent during stable mass transfer when
the radius is inflated. In other words: when the system is a CV, the
mass-transfer rate is too low to be self-regulating. Furthermore, this
parametrization of the mass transfer rate does not correctly model mass
transfer on the thermal time-scale of the donor. We therefore adapted
this factor for both cases separately as described below. 

\subsubsection{Stable CV mass transfer}
To prevent the star from overfilling its Roche lobe by more than a few per
cent, we made the mass transfer prescription more sensitive to changes in the
donor radius by multiplying Eq. (59) from \citet{hurleyetal02-1} with a factor
of 1000 for stable mass transfer when the binary is a CV. 

For a numerically smooth transition, we increased the factor from 1
to 1000 on the same time-scale that we inflated the radius of the donor star. This
allows the mass-transfer rate to be self-regulating during the CV phase and
prevents numerical instabilities in calculating the mass-transfer
rate. 

\subsubsection{Thermal time-scale mass transfer}
In the case of TTMT, the original prescription of \citet{hurleyetal02-1} 
underestimates the mass transfer rate. We therefore included the same factor 
as derived by \citet{claeysetal14-1}, who have tested the resulting calculation of 
the mass transfer rate against a detailed binary stellar evolution code 
(STARS, \citealt{eggleton71-1,polsetal95-1,glebbeeketal08-1}). This provides a better 
description of TTMT, although the mass transfer rate may be a factor of three higher 
than the maximum from the detailed STARS code, in which case the duration of TTMT is 
correspondingly shorter (see their Sect. 2.2.4).

\subsection{Response of the primary to mass transfer}

In the previous sections we have described the consequences of mass transfer
for the mass-losing star. An equally crucial and precarious question is
how the primary star reacts to accretion, in particular, if the mass transfer
is conservative. In what follows, we describe our approach for both TTMT and
mass transfer driven by angular momentum loss.  

\subsubsection{Thermal time-scale mass transfer}
\label{sec:hachisu}
When the mass-transfer rate is within the limits of stable hydrogen burning,
all transferred hydrogen-rich matter is processed into helium and accreted onto
the WD. If the mass-transfer rate exceeds this limit, hydrogen will be
accreted faster than it can be processed into helium. Two scenarios for this
situation have been proposed in the literature. The first is that the
redundant hydrogen-rich matter accumulates onto the surface of the WD and
forms a red-giant-like envelope \citep{nomotoetal79-1}. 
Consequently, the system will probably evolve into a second CE instead of becoming a CV.   
The other scenario is that the burning of hydrogen may cause a very strong
wind \citep{hachisuetal96-1}. This wind ejects part of the accreted matter and
tends to stabilize the mass accretion onto the WD, thus preventing the formation
of a new giant-like envelope. A certain amount of matter will still be
accreted onto the surface of the WD, at a rate $\dot{M}_{\rm acc}$, depending on
the mass accumulation efficiency of hydrogen burning, $\eta_{\rm H}$, and the
mass accumulation efficiency for helium-shell flashes, $\eta_{\rm He}$. This rate
can be expressed as   

\begin{equation}\label{eq:efficiency}
\dot{M}_{\rm acc} = \eta_{\rm H}\,\eta_{\rm He}\,\dot{M}_{\rm tr}
,\end{equation} 
where $\dot{M}_{\rm tr}$ is the rate at which mass is transferred from the
donor to the WD. The efficiency parameters $\eta_{\rm H}$ and $\eta_{\rm He}$
depend on $\dot{M}_{\rm tr}$ \citep{mengetal09-1}. We refer to this wind
as the \emph{Hachisu} wind.  

A second problem concerning the response of the WD to mass accretion occurs if
the WD is made of helium. The accretion of helium can occur either directly or
through stable hydrogen burning on the surface of the WD. Models from
\citet{woosleyetal86-1} have shown that a detonation due to the accretion of
helium at a rate of $2\times 10^{-8} \Msun \mathrm{yr}^{-1}$ can occur when
the star reaches 0.66\,$\Msun$, while previous models by
\citet{nomoto+sugimoto77-1} found a limit of $0.78\,\Msun$ for the accretion
of hydrogen at the same rate. Based on these predictions, the maximum mass of
a helium WD (He WD) was set to $0.7\,\Msun$ in \emph{binary\_c/nucsyn}
\citep{hurleyetal02-1}. However, \citet{shen+bildsten09-1} showed that all of
the pre-1991 results are calculated with an erroneously high value of the
conductive opacity and the resulting ignition masses are a factor of 2 too
low. Furthermore, \citet{saio+nomoto98-1} found that the accretion of helium
onto a $0.4\,\Msun$ WD at a rate of $10^{-7} \Msun \mathrm{yr}^{-1}$ and
$10^{-6} \Msun \mathrm{yr}^{-1}$, that is, the rate of TTMT, induces a stably
burning helium star. This implies that a phase of TTMT can eventually turn a
low-mass He WD into a high-mass carbon-oxygen WD (C/O WD). Since there is
currently no clear stringent limit on the mass of He WDs, if there is any, we
did not assume a mass-limit for accreting He WDs. 

\subsubsection{Mass transfer driven by angular momentum loss}
\label{sec:nova}

When the mass-transfer rate is below the limit of stable hydrogen burning, the
accreted hydrogen is compressed onto the surface of the WD and is subsequently
ignited under highly degenerate conditions. This leads to unstable hydrogen-shell burning and flashes, that is, to nova eruptions. As mass transfer can be
assumed to be stable and conservative in between two (virtually) instantaneous
nova eruptions, the critical mass ratio given by Eq.\,\ref{eq:qad} can be used
to distinguish CVs from systems that are unstable against dynamical mass
transfer, and might evolve into a second CE. However, it has been subject of debate for several
decades how much mass is lost during outbursts. The long-standing paradigm has been that nova eruptions in CVs expel
most of the mass that has been accreted before the nova outbursts
and probably even more \citep{prialnik86-1,prialnik+kovetz95-1,townsley+bildsten04-1,yaronetal05-1}. We
used $m_{\rm acc}$ and $m_{\rm ej}$ from Table 2 in \citet{yaronetal05-1} to 
construct an interpolation table with efficiencies for mass `accretion' during
nova cycles, which we implemented in \emph{binary\_c/nucsyn}. The amount of
mass loss, or in other words, the `accretion' efficiency, depends on the mass of the WD, the
mass-transfer rate, and the core temperature of the WD. We derived the core
temperature of the WD from its mass and luminosity \citep{mestel52-1}. 

As outlined in the introduction, the immense discrepancy between observed and
predicted WD mass distributions in CVs motivated us to 
investigate the implications of WD mass growth during 
nova cycles in a separate model. In these models we arbitrarily 
increased the mass accretion efficiency during a nova cycle from 10 to 100\,\%. 

\section{Binary population synthesis}  
\label{sec:bps}

We generated a three-dimensional grid with \Mone, \Mtwo\, and the separation as
free initial parameters. Our resolution was 150 for each parameter, thus for
each model we simulated $\sim 3 \cdot 10^{6}$ binary systems. We let the
initial mass of the primary, \Mone, range from 1 to 9\,\Msun, to let
the primary evolve into a WD within the Hubble time. The initial mass function
(IMF) of the primary is given by \citet{kroupaetal93-1}. For \Mtwo\, we assumed
an initial 
mass-ratio distribution that is flat in mass 
ratio $q$ \citep{sanaetal09-1}. 
The initial mass of \Mtwo\, ranged from 0.08 to 3.5\,\Msun. Both \Mone\, and
\Mtwo\, were picked from their initial mass distribution with a logarithmic
spacing. The initial orbital separation $a$ was assumed to be flat in log $a$
\citep{popovaetal82-1, kouwenhovenetal07-1} and ranged from 3 to $10^{4}$
\Rsun\, to cover the whole space of binaries that will interact within the
Hubble time. Each binary in the grid was given an individual formation
probability, depending on the assumed distribution of the initial parameters
and taking into account the logarithmic spacing in mass. We furthermore assumed
circular orbits and solar metallicity for all binaries. We set the CE
efficiency parameter, $\alpha_{\rm ce}$, equal to 0.25, in accordance with the
range of values determined by \citet{zorotovicetal10-1}. 
For details of the CE prescription we used, see 
\citet{hurleyetal02-1} and \citet{zorotovicetal10-1}.
The binaries were
formed with a constant star formation rate by randomly assigning them a
lifetime between 0 and 13.5 Gyr, our assumed age of the Galaxy
\citep{pasquinietal04-1}.  

The motivation for this work is the disagreement between observed and
predicted WD mass distributions of CVs. \citet{zorotovicetal11-1} suggested
two possible solutions for the high WD masses in the observed sample, and we
calculated three models to test both ideas against the standard scenario of CV
evolution. In the first model, we assumed that a giant-like envelope is formed
when the mass transfer rate exceeds the limit for stable hydrogen burning,
which should produce relatively few post-TTMT CVs. This model, which we
refer to as the reference model, contains the most common assumptions of CV
formation and evolution as discussed in the previous sections. In our second
model, we added the stabilizing \emph{Hachisu} wind during TTMT as this is known
to be fundamental to produce large numbers of post-TTMT CVs, which potentially
contain massive WDs. In a third model, we incorporated net mass growth during
nova cycles into our reference model instead of the `accretion' efficiencies
derived from \citet{yaronetal05-1}.  

The three main models can thus be summarized as follows:
\begin{enumerate}
\item Our reference model based on typical assumptions for CV evolution. 
\item Our reference model, including a wind from the accreting WD that stabilizes mass 
transfer and prevents the formation of a giant-like envelope (Sect. \ref{sec:hachisu}).
\item Our reference model, but assuming different efficiencies of mass accretion onto the WD.  
\end{enumerate}

\section{Results}

We first describe the results of our reference model for CV
evolution. Subsequently, we address the effects of including 
the \emph{Hachisu} wind, and then the mass growth.  

\subsection{Reference model} 

\begin{figure*}[!ht]
  \begin{center}
    \includegraphics[width=0.99\textwidth]{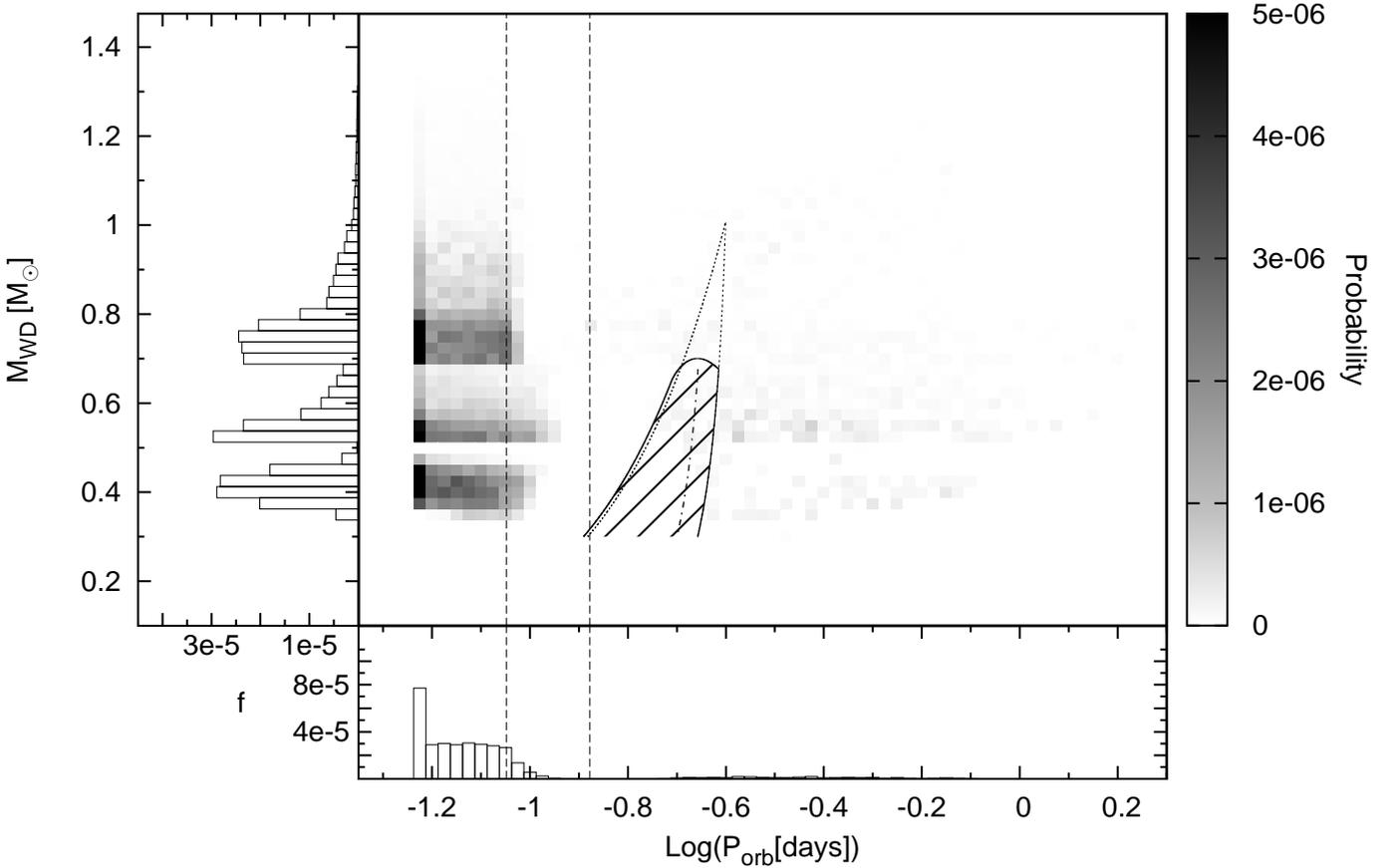}
    \caption{\emph{Centre}: Two-dimensional histogram of the orbital period and 
    WD mass distributions in CVs for our reference model. The colour intensity 
    represents the sum of the formation probabilities of all CVs residing in that
    two-dimensional bin. The vertical dashed lines mark the upper (3.18 hr) and lower 
    edge (2.15 hr) of the observed period gap \citep{knigge06-1}. The hatched region 
    indicates the parameter space in which mass transfer will become dynamically unstable, 
    assuming $q_{\rm cr}$ is given as in Eq. (\ref{eq:qad}). The dotted line marks the shape
    of the dynamically unstable region if one assumes a constant $q_{\rm cr}$ of $0.695$ for
    $\Mtwo \lesssim 0.7\,\Msun$. These regions are depicted in the same way as their 
    corresponding regions in Fig.\,\ref{fig:webbink}. The dashed-dotted line marks the 
    region where mass transfer becomes dynamically unstable if one assumes the ZAMS 
    mass-radius from \citet{toutetal96-1}. \emph{Bottom panel}: orbital period distribution 
    of the CVs in our reference model. \emph{Left panel}: WD mass distribution of the CVs 
    in our reference model. \label{fig:refmodel}} 
  \end{center}
\end{figure*}

Figure\,\ref{fig:refmodel} shows the binaries that are identified as a CV at
the present epoch (according to our definition of a CV, see Sect.
\ref{sec:stab_mt}). The distribution of both the orbital period (bottom) and
WD mass (left) are shown, as well as their combined probability distribution
in the two-dimensional plane. The vertical dashed lines mark the upper (3.18 hr)
and lower edge (2.15 hr) of the observed period gap \citep{knigge06-1}. The
hatched region indicates the parameter space in which mass transfer will
become dynamically unstable, assuming $q_{\rm cr}$ as given as in
Eq. (\ref{eq:qad}). The dotted line marks the shape of the dynamically
unstable region if one assumes a constant $q_{\rm cr}$ of $0.695$ for $\Mtwo
\lesssim 0.7\,\Msun$. These regions are depicted in the same way as their
corresponding regions in Fig.\,\ref{fig:webbink} and were calculated by
equating the radius of the donor star, as given by \citet{kniggeetal11-1},
with its Roche-lobe radius and solving for the orbit with Kepler’s third law
using the corresponding critical mass ratio. The right border of this region
corresponds to a donor mass of 0.7\,\Msun, the curved left border
corresponds to the critical mass ratio. Some CVs are located in the region for
dynamically unstable mass transfer. To explain this, we have also shown the
right border for this region if one assumes the mass-radius relation that is
originally in the code \citep{toutetal96-1}, indicated by the dashed-dotted
line. Since virtually all of the CVs in the hatched region are located on the stable
side of the dashed-dotted border, it illustrates that these CVs have 
formed recently and their radius is still inflating towards the value given by
\citet{kniggeetal11-1}. The mass of the CV donors 
in this region is higher than $0.7\,\Msun$ 
and the system is currently still stable against mass transfer. 
These CVs, however,  
will experience 
dynamically unstable mass transfer when the donor star becomes
deeply convective at a donor mass of $\sim0.7\,\Msun$.  

The orbital period distribution shows a clear spike at the period minimum,
corresponding to very low-mass secondaries ($\Msec \lesssim
0.08\,\Msun$). Most CVs in our standard model are
currently below the period gap, but only 33.6\,\% of the current CV
population was born in or below the gap. This is a well-known prediction of
binary population models of CVs \citep[e.g.][]{kolb93-1}: 
because angular momentum loss
is assumed to be much lower below the gap than above the gap, the
evolutionary time-scale is shorter above the gap than below by one to two orders
of magnitude. 

The absence of CVs with low-mass WDs in the $\sim\,3-4$\,hr orbital period
range, that is, in the hatched region in Fig.\,\ref{fig:refmodel}, has been predicted
by previous models \citep{dekool92-1} and can be explained as follows: mass
transfer becomes dynamically unstable for CVs whose mass ratio exceeds the
critical value given by Eq. (\ref{eq:qad}). This will occur within the region
marked by the solid line in Fig.\,\ref{fig:refmodel}. CVs are thus not able to
evolve towards shorter periods through this region without experiencing
dynamically unstable mass transfer and merging. This means CVs with a WD
$\lesssim 0.7\,\Msun$ and $log(\frac{P_{\rm orb}}{\mathrm{days}}) \lesssim
-0.7$ have to be born there or must have descended from higher WD masses.  

The WD mass distribution shows three peaks. The first peak at
$0.4\,\Msun$ consists of He WDs that evolved into a CE when the primary was on 
the first giant branch (FGB). The second peak at 0.55\,\Msun\, corresponds
to binaries for which the mass growth of their core was
terminated when they were on the asymptotic giant branch (AGB). These are C/O WDs 
with a lower mass than expected for single WDs as a result of the earlier expulsion of the envelope. 
The steep decline after the peak for C/O WDs agrees with the observed mass distribution
of single WDs \citep{kepleretal07-1}. The third peak predicted at
$0.75\,\Msun$ is atypical for the mass distribution of single WDs. This peak
contains a significant fraction (32.5\,\%) of CVs that have experienced TTMT and thus
originates from CVs that initially resided at longer orbital periods and that
had more massive secondaries. 
It is a direct consequence of using Eq. (\ref{eq:qad}) for the critical 
mass ratio at which mass transfer becomes dynamically unstable instead of 
a constant value for $q_{\rm cr}$, as is
assumed by default in \emph{binary\_c/nucsyn}. 
The region in which mass transfer becomes dynamically unstable
for a constant $q_{\rm cr}$ of $0.695$ for $\Mtwo \lesssim 0.7\,\Msun$ extends
towards high-mass WDs as indicated by the dotted line in
Fig.\,\ref{fig:refmodel}. Thus, in this case, all CVs with a WD $\lesssim
1\,\Msun$ and $log(\frac{P_{\rm orb}}{\mathrm{days}}) \gtrsim -0.6$ would
merge eventually. The occurrence of this third peak has been previously predicted by CV
population models, using the same prescription for the critical mass ratio \citep[see][ his Fig.\,4]{politano96-1}. 
If this prescription had not been used, the predicted WD mass distribution
would be similar to the one derived by \citet{dekool92-1}.   

Our reference model predicts that 15\,\% of the CVs have undergone TTMT
before evolving into a CV, which is consistent with the results from UV observations 
from \citet{gaensickeetal03-1}. 
The relatively small fraction of such post
TTMT CVs is caused by the formation of a giant-like envelope if the
mass-transfer rate of hydrogen-rich matter exceeds the limit for stable
hydrogen burning. Consequently, He WD primaries evolve into FGB stars and C/O
or O/Ne WDs primaries into AGB stars. Therefore, the binaries do not evolve
into CVs, but eventually evolve into a second CE and merge. In other words,
the limits for the mass-transfer rate, between which hydrogen burning on the
surface of the WD is stable, define a range that is very narrow. Mass-transfer
rates are thus typically either too high, causing the CVs to merge, or too low
to allow the WD to gain a sufficient amount of mass. 
This limits the impact of post TTMT CVs on the WD mass distribution of CVs in
our reference model. 

\subsection{CV populations including the \emph{Hachisu} wind} 

\begin{figure*}
  \begin{center}
    \includegraphics[width=0.99\textwidth]{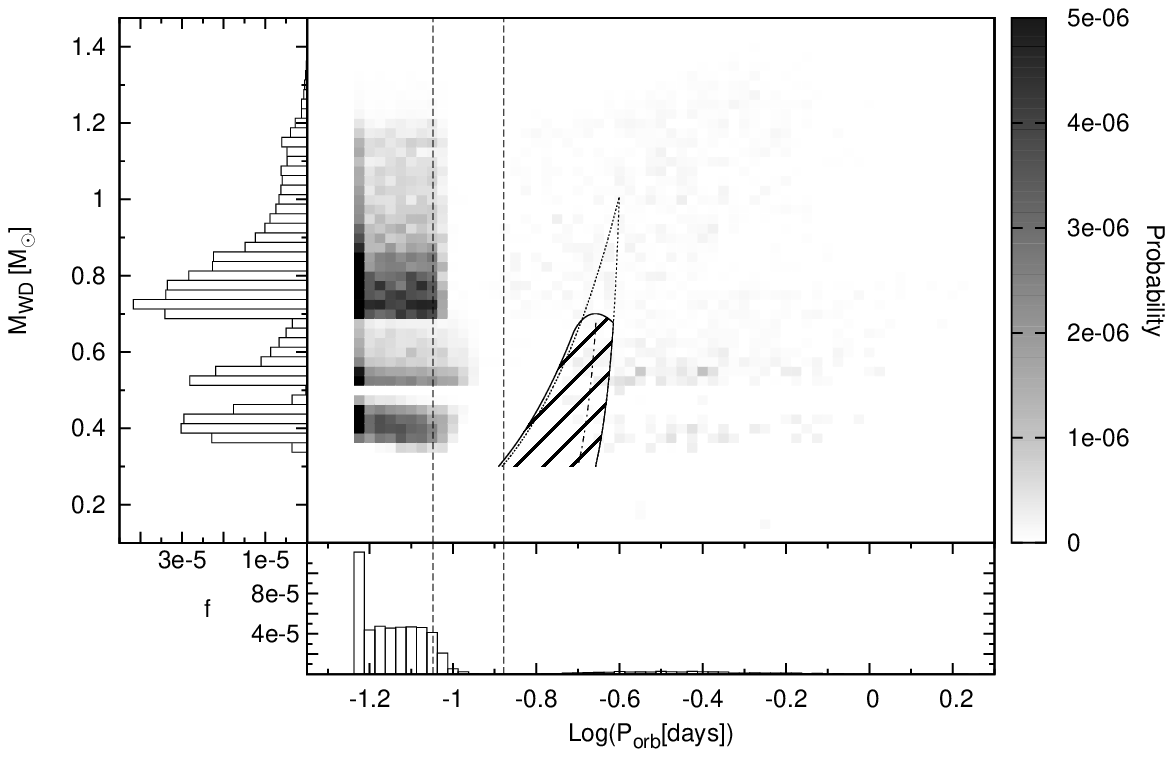}
    \caption{Same as in Fig.\,\ref{fig:refmodel}, but for model 2, in which a wind 
    from the WD that stabilizes mass transfer is included (Sect. \ref{sec:hachisu}). 
    This wind allows more binaries that experience TTMT to evolve into CVs. The prescription 
    for $q_{\rm cr}$ is crucial for the subsequent CV evolution.\label{fig:hachisu}}
  \end{center}
\end{figure*}

In our second model the implementation of the \emph{Hachisu} wind (see
Sect.\,\ref{sec:hachisu}) prevents the WDs from forming a new giant-like
envelope in the case of TTMT. This wind thus serves as a tool for investigating
the consequences of extending the narrow range for stable hydrogen burning to
higher mass-transfer rates, that is to say, extending the parameter space for TTMT
systems that may evolve into CVs.  
The resulting WD mass (left) and orbital period
(bottom)  distributions are shown in Fig.\,\ref{fig:hachisu}. As in
Fig.\,\ref{fig:refmodel}, the vertical dashed lines indicate the location of
the observed orbital period gap, while the solid and dashed lines in the main
grey-scale plot represent the region in which mass transfer becomes
dynamically unstable according to the two prescriptions for $q_{\rm{cr}}$
discussed previously. The dashed-dotted line marks the dynamically unstable
region, assuming the mass-radius relation from \citet{toutetal96-1}.  

The WD mass distribution for $\Mwd \lesssim 0.7\,\Msun$ is practically the
same as the distribution predicted by the reference model shown in
Fig.\,\ref{fig:refmodel}, and so are their evolutionary paths. Assuming a
wider range of mass-transfer rates that allow stable hydrogen burning without
the formation of a new giant-like envelope, however, leads to the prediction
of a much larger number of CVs with massive WDs ($\gtrsim 0.7\,\Msun$), which
in the reference model evolved into a giant-like star. This increases
the total number of predicted CVs by 50\,\% with respect to the reference
model and now almost half of all CVs,  46\,\%, have undergone TTMT. The 
mean WD mass of post-TTMT CVs increased significantly during TTMT: from
$0.56\,\Msun$ before the phase of TTMT, to $0.86\,\Msun$ afterwards. When
only the CVs with a WD $\gtrsim 0.7\,\Msun$ are considered, 74.8\,\%
experienced TTMT.  

The impact of the description used for $q_{\rm cr}$ on the predicted
distributions is even stronger than in the reference model. Most WDs with masses
between $0.7\,\Msun$ and $1\,\Msun$ would not evolve into CVs if one assumes a
constant $q_{\rm cr}$ of $0.695$, because they would experience dynamically
unstable mass transfer when they evolve towards the region bordered by the
dotted line (see Fig.\,\ref{fig:hachisu}). In that case, the only possibility
to form WDs in CVs with a mass around $0.8\,\Msun$ below the gap are initial
primary masses $\gtrsim 3\,\Msun$, which have a very low formation
probability compared to less massive stars.   

The predicted WD mass distribution is fairly broad above the gap and
more strongly peaked below, while the mean WD mass is comparable for
both distributions (0.76 and 0.70\,\Msun). This is caused 
by two effects. First, as most CVs with
high-mass WDs below the gap descend from CVs above the gap (meaning
that they are not born
below the gap), the assumed mass loss during nova cycles as described in
Sect.\,\ref{sec:nova} slightly erodes these massive WDs ($0.8-1.2\,\Msun$) on
their way from long ($\sim$ 9.5 hours, i.e. $log(\frac{P_{\rm
    orb}}{\mathrm{days}}) \sim -0.4$) to short orbital periods. Second, CVs
containing low-mass WDs ($0.4-0.6\,\Msun$) above the gap cannot avoid
dynamically unstable mass transfer when evolving towards shorter periods,
that is, they enter the hatched region in Figs.\,\ref{fig:webbink}
and\,\ref{fig:hachisu} when the secondary develops a deep convective envelope
(at a secondary mass of $\sim0.7\,\Msun$) and merge during a second
CE. Consequently, the distribution of WD masses above the gap shows a wider
spread than the distribution of WD masses below the gap.  

\subsection{Mass growth during nova cycles}  

\begin{figure}
   \centering
    \includegraphics[width=0.49\textwidth]{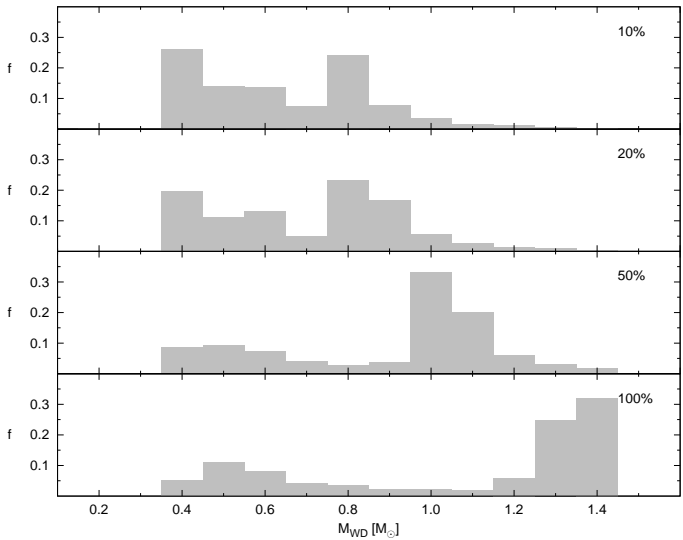}
    \caption{WD mass distributions for different fractions of mass growth 
    during nova cycles, i.e. assuming that more mass is accreted between two nova
    outbursts than is expelled during the eruption. The fractions are normalized 
    to the total formation probability of all the CVs in each model. A higher 
    accretion efficiency shifts the peak at $0.8\,\Msun$ to higher masses.
\label{fig:ma-gr}}
\end{figure}

In the third model, we investigate the other possibility to produce many CVs containing high-mass WDs mentioned by
\citet{zorotovicetal11-1}: we tested whether net mass growth of the WD in 
a CV evolving through a nova cycle can shift the low-mass WDs towards higher
masses. 
We used the same assumptions as in our reference model, but instead of the
standard assumption of slow erosion of the WD, we here tested four different
accretion efficiencies. Figure\,\ref{fig:ma-gr} shows the WD mass
distributions assuming that 10, 20, 50, or 100\,\% (from top to bottom) of the
transferred mass is accreted, which means that it remains on the WD after a nova
outburst. These WD mass distributions are normalized by the sum of the
formation probabilities of all CVs in each model. Assuming an accretion
efficiency of only 10\,\% has little consequences for the predicted WD
masses. The mean mass increases slightly, but the overall shape of the
distribution with its three peaks remains the same (the peak at 0.55$\,\Msun$
is smoothed out because of the bin size). This is simply because the 
low-mass WDs with relatively high-mass secondaries still evolve into the
dynamically unstable mass transfer region and those with low-mass secondaries
gain very little mass. The situation remains relatively similar for an
accretion efficiency of 20\,\%, but changes significantly if 50\,\% of the
transferred mass is assumed to contribute to mass growth of the WD. In the
latter case, many CVs with low-mass WDs and initially massive secondaries can
circumnavigate the dynamically unstable region and evolve into CVs below the
gap. In addition, many WDs gain a significant amount of mass. As a consequence,
the peak located at $0.75\,\Msun$ in the distribution predicted by the
reference model largely dominates the distribution and is moved towards higher
masses ($\sim 1.0\,\Msun$). 
This effect becomes stronger if all the transferred mass is assumed to
be accreted by the WD. In this case, many CVs evolve into SN\,Ia candidates
with WD masses $\gappr\,1.2\,\Msun$. The latter model is, of course, not a
realistic one (because a significant fraction of mass is obviously expelled during
nova eruptions), but it serves to illustrate the potential effect of mass
growth on the WD mass distribution in CVs.  

\section{Discussion} 

Table\,\ref{tb:summary} shows the statistics for our different simulations.
Our results can be summarized as follows: 
the reference model predicts the WD mass distribution to be dominated by
low-mass WDs and a third peak at $\sim 0.75\,\Msun$ where most post-TTMT CVs reside.
The fraction of CVs that experienced a previous phase of
TTMT increased by a factor of three when we incorporated the 
\emph{Hachisu} wind, which makes the peak at 
$\sim 0.75\,\Msun$ more prominent. 
Finally, the models in which we allowed for mass growth during the CV phase
predict relatively small changes for low accretion efficiencies, that is, a slight 
increase of the number of systems containing high-mass WDs. High fractions of 
mass growth predict CV populations dominated 
by systems containing very massive WDs ($\gappr1.0\Msun$). 

\begin{table*}[!ht]
\begin{center}
\caption{Statistics of our different models. \emph{Columns 2 to 6}: the fraction of binaries in the grid that 
currently are a CV, the average WD mass of the CV population, the fraction 
of CVs that have a central hydrogen fraction $<$ 0.4, the average WD mass at the
beginning and at the end of the TTMT phase and 
the percentage of CVs that had a phase of TTMT. Assuming a `stabilizing' wind from
the WD increases the current CV population by 50\,\% and triples the number of
TTMT descendants. On the other hand, increasing mass growth allows more CVs that 
emerge from a phase of TTMT with a (on average) lower WD mass to circumnavigate
dynamically unstable mass transfer.}
{\begin{tabular}{|l|c|c|c|ccc|}
\hline
&&&&\multicolumn{3}{c|}{\textbf{CVs With TTMT}}\\
\cline{5-7}
\textbf{Model}&\textbf{CVs In Grid}&$\mathbf{\langle M_{\rm \textbf{WD}}
  \rangle}$&\textbf{Evolved Donors}&$\mathbf{\langle M_{\rm \textbf{WD,prior}}
  \rangle}$&$\mathbf{\langle M_{\rm \textbf{WD,after}} \rangle}$&\textbf{\%}\\

& (\%)&($\Msun$)& (\%)&($\Msun$)&($\Msun$)&\\
\hline
Reference&0.40&0.62&23.38&0.51&0.82&15.0\\
\emph{Hachisu} wind&0.63&0.71&37.22&0.56&0.86&46.0\\
\hline
Mass growth&&&&&&\\
\multicolumn{1}{|r|}{10\,\%}&0.40&0.65&19.96&0.54&0.93&7.3\\
\multicolumn{1}{|r|}{20\,\%}&0.47&0.71&24.82&0.51&0.96&8.8\\
\multicolumn{1}{|r|}{50\,\%}&0.70&0.90&35.41&0.48&1.09&9.1\\
\multicolumn{1}{|r|}{100\,\%}&0.63&1.08&29.96&0.48&1.34&11.2\\
\hline
\end{tabular}}
\label{tb:summary}
\end{center}
\end{table*}

Below we first address to what extend 
our results depend on the assumptions and simplifications we made on our
modelling approach, before we compare our model predictions with the 
observations.
 
\subsection{Model uncertainties}
The model predictions are sensitive to several assumptions, for
instance to the critical
mass ratio for dynamically unstable mass transfer, the \emph{Hachisu} wind,
the calculation of the TTMT rate, the mass limit for He WDs, the initial 
mass-ratio distribution, the strength of MB, and the efficiency of CE evolution.  

\subsubsection{Critical mass ratio}

We demonstrated that the simulation of CV evolution is extremely sensitive to
the analytic fit for $q_{\rm cr}$ from \citet{politano96-1}. This analytic fit
allows a significant number of CVs with a massive WD to evolve towards shorter
periods, while they would have been regarded as  dynamically unstable if one
would have used a constant $q_{\rm cr}$ for all low-mass MS stars, that is, a
rough cut-off at $\sim 0.7\,\Msun$. The critical mass ratio, and the
corresponding region in which mass transfer becomes dynamically unstable,
functions as a kind of `road block' that prevents CVs with low-mass WDs to
evolve towards shorter orbital periods. Instead, dynamically unstable mass
transfer leads most likely to a second CE and the stars probably merge. 
The value of $q_{\rm cr}$ therefore has a huge
influence on the formation of $\sim\,0.8\,\Msun$ WDs. Even fine-tuning all
other parameters towards producing many post TTMT CVs, such as the
\emph{Hachisu} wind, would predict few WDs around $0.8\,\Msun$ if $q_{\rm cr}$ were 
constant. It is therefore of crucial importance to use an accurate value of $q_{\rm cr}$
instead of a crude approximation.  

Since the critical mass ratio depends on the adiabatic mass-radius exponent,
it is mainly determined by the stellar structure of the donor star and whether
mass transfer is conservative or not. To our knowledge, the prescription we
use for $q_{\mathrm{cr}}$ is the most accurate currently available, but it is
still based on two assumptions that might be critical for the formation of
CVs, that is, a MS structure of the donor and
conservative mass transfer. 

First, the detailed structure of the secondary stars evolving from TTMT,
especially the mass at which a convective envelope develops, is quite
uncertain because these systems were initially more massive. These donors are
probably driven out of thermal equilibrium by mass transfer at a high rate
and their structure might thus to some degree correspond to that of a more
massive star. If this is the case, the mass at which a deep convective
envelope develops, that is, where the adiabatic mass-radius exponent increases
steeply, 
could be lower than the canonical value of $0.7\,\Msun$. Thus, for a given
WD mass, the mass ratio of a system evolving from long orbital periods to
shorter orbital periods could be lower when facing the dynamically unstable 
boundary. Therefore more CVs with less
massive WDs ($\lappr\,0.75\,\Msun$) would evolve towards shorter periods 
without
merging because the left-shifted boundary provides a higher $q_{\rm cr}$ 
for a given secondary mass. 
This would concern CVs born at long orbital
periods and post-TTMT systems. 
The peak of the WD mass distribution at $0.75\,\Msun$ in the
reference model (and in the model including the \emph{Hachisu} wind)
would probably be shifted 
towards slightly lower masses. The impact of shifting the mass limit at
which the donor stars develop a deep convective envelope towards lower
masses on the resulting WD mass distribution would be even stronger when the
masses of the WDs are assumed to grow during the CV phase, which would allow
low-mass WDs at long orbital periods to circumnavigate dynamically unstable
mass transfer if they accrete a sufficient amount of mass. 

Second, we assumed that stable mass transfer (driven by angular momentum loss)
during the CV phase is conservative. 
However, CVs experience nova eruptions during which both mass and
  angular momentum is lost. 
If the angular momentum taken away by the mass expelled during a nova
outburst significantly exceeds the specific angular momentum of WD material,  
the critical mass ratio would be lower for a given value of the
mass-radius exponent than it would be in the case of conservative mass
transfer. This would
cause stable mass transfer to occur only in systems with low mass ratios 
and may solve the problem of the large number of predicted CVs with low-mass
WD. We will discuss this possibility in a follow-up paper.

This reasoning and the sensitivity of the results to the critical mass 
ratio imply that we may underestimate 
the relative number of massive WDs in our models. 
The adiabatic mass-radius exponent as shown in
Fig.\,\ref{fig:webbink} is currently being scrutinized (Nelemans, Webbink, private
communication). A more accurate value for mass-transferring stars based on
detailed stellar models would drastically reduce the uncertainties in our
model predictions. 

\subsubsection{\emph{Hachisu} wind}

Since there is a consensus on the regime in which the mass-transfer rate enables 
stable hydrogen burning \citep{nomotoetal07-1,williams13-1}, the only option to 
increase the number of CVs that descend from TTMT is to assume non-conservative mass 
transfer. The most popular mechanism for mass loss during TTMT is a strong wind that 
reduces the mass accretion rate and allows more systems to maintain stable hydrogen 
burning on the surface of the WD \citep{hachisuetal96-1}. In the framework of this wind
model, the mass accumulation efficiency for hydrogen burning, $\eta_{\rm H}$, and in 
particular the mass accumulation efficiency for helium shell flashes, $\eta_{\rm He}$, 
during the wind determine how much the WD is growing and thus how fast the mass ratio 
changes as the mass of the secondary decreases. These efficiencies therefore define the
evolutionary course along which a system evolves towards lower secondary masses, that is, to 
the left in Fig.\,\ref{fig:webbink}. Higher efficiencies would imply a steeper decline of
the mass ratio and more CVs with a phase of TTMT could circumnavigate dynamically unstable 
mass transfer. In contrast, lower efficiencies imply that CVs with a phase of TTMT are more 
likely to run into dynamically unstable mass transfer. Thus, the relative number of post-TTMT
CVs, and consequently also the resulting WD mass distribution, depend sensitively on the mass 
accumulation efficiencies during TTMT. 

\subsubsection{Calculating the thermal time-scale mass transfer rate}
The calculation of the mass transfer rate in \emph{binary\_c/nucsyn} does not
depend on the mass-radius exponent. In the case of TTMT
the prescription from \citet{claeysetal14-1} may overestimate the
mass transfer rate. We therefore also ran our simulations with the
original prescription, which underestimates the mass transfer rate. The
outcomes differ in which binaries actually become a CV after TTMT, but the
overall distributions of both the WD mass and the orbital period look the
same. We therefore argue that using a method that models TTMT more adequately
\citep[see e.g.][]{chenetal14-1} would not (significantly) change the WD mass distribution. The period
distribution, on the other hand, may look different since we have a large
number of evolved donor stars in our CV population. We address this 
question below (see also Sect.\,\ref{sec:evolved_donors}).


\subsubsection{Accretion onto He WDs}
\label{sec:discussmtr}
An important uncertainty that has not been tested by our model calculations is
the maximum mass of He WDs that accrete hydrogen rich material (that might be
fused into He on the surface of the WD). According to
the literature, this value is highly uncertain, as outlined in
Sect.\,\ref{sec:hachisu}. If He WDs can be more massive than $0.7\,\Msun$,
they provide a serious contribution to the observed peak at $0.8\,\Msun$ and
simultaneously reduce the relative number of WDs $\lesssim 0.5\,\Msun$. 
In the model including the \emph{Hachisu} wind (see Fig.\,\,\ref{fig:hachisu}) 
the contribution of He WDs that accreted a significant amount of mass during 
TTMT to the peak at $\sim0.8\,\Msun$ reaches 24.9\,\%.  
Thus, if there exists a mass limit for accreting He WDs, the
predicted peak in the WD mass distribution 
will be significantly less pronounced.

\subsubsection{Initial mass-ratio distribution}

One of the most important and critical ingredients of binary population models
is the initial mass-ratio distribution. We assumed that all initial 
mass ratios are equally probable. Although both early \citep[][]{popovaetal82-1} as
well as more recent observational works \citep{raghavanetal10-1} 
seem to support a mass-ratio distribution favouring equal-mass binaries, the general consensus is that the initial mass ratio distribution is flat, 
even for massive stars that appear to be most biased towards equal 
masses \citep{sanaetal09-1, sanaetal12-1}. 
An initial mass-ratio that is flat in q does not favour the formation of
nearly equal-mass binaries, which is essential to produce large fractions of
post-TTMT CVs. An initial 
mass-ratio distribution that is proportional to $q$ would thus significantly
increase the fraction of CVs containing high-mass WDs in the resulting
distributions. This would cause the peak at 0.75\,\Msun \ to dominate the
peaks at 0.55 and 0.4\,\Msun\ more strongly, as is observed.

\subsubsection{MB and CE efficiency}
\label{sec:mbce}
We have assumed MB to be inefficient for fully convective secondary
stars. This, so named, disrupted MB scenario has been well
established during the past decade both in single stars \citep[see e.g.][ and
  references therein]{reiners+mohanty12-1} and in close compact binary stars
\citep{schreiberetal10-1}. However, the strength of MB when the secondary star
contains a radiative core is not well known. Current prescriptions differ by
several orders of magnitude
\citep{schreiber+gaensicke03-1,kniggeetal11-1}. Reducing the strength of MB, for instance by considering the normalization factor implemented by
\citet{davisetal08-1}, would increase the evolutionary time-scale of detached
PCEBs with secondary stars in the mass range $\sim0.35-0.7\,\Msun$ and
slows CV evolution towards shorter periods for systems with donor star
masses in the range $\sim0.2-0.7\,\Msun$. This implies that CVs with high-mass
WDs, that is, those systems that form the peak at $\sim0.8\,\Msun$ in the model
including the \emph{Hachisu} wind, would
dominate the systems that become CVs below the 
period gap less strongly at shorter periods. 
Furthermore, owing to the slower 
evolution towards shorter orbital periods of CVs above the gap, our models
would predict more systems in this period range for decreased MB. 

Equally uncertain as the strength of MB, but perhaps even more important for
close compact binary evolution, is the efficiency of CE evolution. It seems
that, at least for low-mass secondary stars ($\sim0.1-0.5\,\Msun$), a
relatively low value of the CE efficiency ($\sim0.25$) agrees best with the
observations \citep{zorotovicetal10-1} and numerical simulations
\citep{ricker+taam12-1}. However, CE evolution is extremely uncertain mostly because
of three reasons. First, we have currently no idea if this low value
also holds for higher secondary masses, second, we do not know if perhaps a
post-CE circumbinary disk efficiently extracts angular momentum from the
binary orbit of the emerging detached PCEB \citep{kashi+soker11-1}, and
finally, it also remains an open discussion if perhaps the internal energy
stored in the envelope such as the recombination energy contributes to
the envelope ejection process \citep{webbink08-1,rebassa-mansergasetal12-1}. 

However, the main effect that the CE efficiency would have on the predicted WD
mass distributions is that higher values lead to the prediction of more CVs
containing He WDs. This is simply because a significantly larger fraction of
initially close binaries that start mass transfer when the primary star is on
the first giant branch can survive CE evolution. This has recently been
confirmed by \citet{zorotovicetal14-1}, who showed that increasing $\alpha_{\rm ce}$ 
from 0.25 to 1 results in an increasing fraction of PCEBs with He WDs and an 
extension of the WD mass distribution towards lower masses that cannot be 
obtained with lower efficiencies. Therefore, using a higher value of $\alpha_{\rm ce}$ 
would increase the disagreement between our simulations and observations.

\subsection{Comparison with observations}

The model uncertainties presented in the previous section imply that
performing binary population models of CVs has to be based on several
assumptions, and changing these assumptions affects the predicted 
distributions. We have shown that mass growth during the CV phase 
can shift large fractions to higher masses but struggles to predict a 
broad distribution with the mean value around $0.8\,\Msun$. We have also shown 
that the latter can naturally be obtained if a large fraction of CVs 
are assumed to descend from TTMT, but a large population of CVs with He WDs remains in all our models compared to the observations. 
Below we compare in detail the model predictions with the observations 
of CVs and their progenitors.  
Because the orbital period is an observable that is much easier to measure
than the stellar masses, we first inspect whether the orbital period distributions 
predicted by our models agree
with the observations before we take a detailed
look at the predicted WD mass distributions. 

\subsubsection{Orbital period distribution}

In general, our simulations predict orbital period distributions that resemble
those of previous models \citep{kolb93-1,howelletal01-1} 
and agree well with the main features of the observed distribution. 
The predicted period minimum, the accumulation of systems at the period minimum, the orbital period gap, 
and the prediction of most CVs residing below the gap all agree with the largest homogeneous 
sample of observed CVs currently available \citep{gaensickeetal09-1}. This agreement is expected as the predicted period minimum and period gap are a direct consequence of the mass-radius 
relation we used for CV donor stars \citep{kniggeetal11-1}, which has been designed to reproduce the 
period gap and the period minimum. 

However, our models disagree with the observed distributions with respect to
some details. For example, we predict a deficit of CVs between
$log(\frac{P_{\rm orb}}{\mathrm{days}}) = -0.9$ and $-0.7$ but a
number of CVs at longer orbital periods, while the opposite seems to be
observed \citep{gaensickeetal09-1}. Furthermore, the accumulation of systems
at the orbital period minimum seems to be more pronounced in our simulations
than in the observed sample. These disagreements are caused by both model
uncertainties as described in the previous section and, probably more
important, observational biases and selection effects. For instance, while the
predicted relative number of CVs above the gap sensitively depends on the
strength of MB and the CE efficiency, the disagreement between the predicted
number of these systems and the observations is most likely dominated by
observational biases. CVs at the upper edge of the period gap are particularly
bright (high mass-transfer rates) while systems at the period minimum are
particularly faint (low mass-transfer rates). Both the larger number of
systems at the upper edge of the gap and the less pronounced peak at the
orbital period minimum are therefore probably caused by the fact that bright
systems are more likely to be observed. This becomes clear when comparing the orbital period distribution of the 
complete sample of CVs with the most homogeneous sample identified so far 
\citep[see][ their Fig.\,2]{gaensickeetal09-1}. 
Similar results are obtained if observational biases are taken into account. In fact, 
\citet{pretoriusetal07-1} have shown that observational biases impose a large discrepancy 
between the predicted and observed period distribution of CVs. Their predicted period distribution 
looks very similar to the period distribution of our models (see their Fig.\,11).

\subsubsection{WD mass distributions of PCEBs}

\begin{figure}
    \centering
    \includegraphics[width=0.49\textwidth]{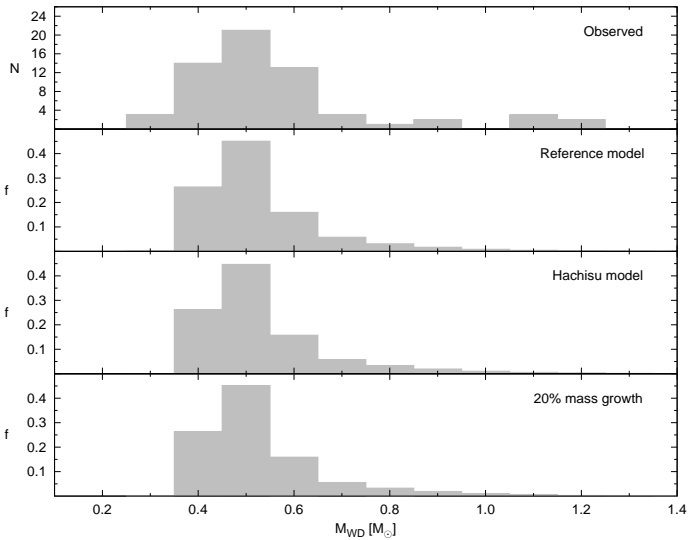}
    \caption{Mass distribution of WDs in PCEBs. \emph{From top to bottom}:
      observed sample used in \citet{zorotovicetal11-1}, distribution
      predicted by our reference model, distribution derived from the
      reference model including a strong wind during TTMT, distribution
      predicted if it is assumed that 20\% of the transferred mass during a
      nova cycle remains on the WD. The fractions are normalized to the total
      formation probability of all PCEBs in each model. The predicted 
PCEB distributions are identical (because mass growth during TTMT or 
during nova cycles does not affect the WD masses of the progenitors)  
and agree reasonably well with the observations. 
\label{fig:pceb}}
\end{figure}

As shown by \citet{zorotovicetal11-1}, the high WD masses in CVs are not
imprinted on the observed sample of PCEBs containing low-mass (M-dwarf)
secondary stars. This led the authors to the conclusion that either the
currently known 
pre-CV sample is not representative for CV progenitors or the masses of the
WDs in CVs grow during nova cycles. In this paper we tested both
possibilities. Before we directly compare the predicted WD mass distributions
for CVs, we first test whether the PCEB WD mass distributions of our models
agree with the observed one. The corresponding distributions are shown in
Fig.\,\ref{fig:pceb}, where we define a binary as a PCEB if it had a CE,
currently does not have Roche-lobe overflow, and consists of a WD (He, C/O,
O/Ne) and a MS star. We
used relatively large bins because the uncertainties in the masses
derived from observations are significant. 
As expected, the WD mass distribution of PCEBs
does not depend on our assumptions because both mass growth during nova cycles or
during a phase of TTMT occur after the detached PCEB phase. 
Furthermore, the simulated WD mass distribution of PCEBs
looks similar to the distribution of the PCEB
sample from \citet{zorotovicetal11-1}. Both distributions have the same
tendency to peak at $0.5\,\Msun$ and show similar scatter around their mean
value. However, our models fail to reproduce the extremely high-mass WDs
($\gappr0.8\,\Msun$) that seem to be present in the observed
distributions. This is probably due to uncertainties in the masses derived
from observations. The WD masses shown in the top panel of
Fig.\,\ref{fig:pceb} have been measured using the spectral
decomposition/fitting method described in
\citet{rebassa-mansergasetal07-1}. We recently learned that more robust
measurements of the WD mass using eclipse light curves indicate a much lower
mass for those systems were the spectra indicate massive WDs \citep[see][ for
  one example]{parsonsetal13-1}. Therefore, the disagreement
between prediction and observation concerning the massive WDs in PCEBs is most
likely caused by limitations of the spectral decomposition/fitting method. We
also conclude from Fig.\,\ref{fig:pceb} that, according to all three types
of models, either virtually all CVs with a WD more massive than $\sim0.8\,\Msun$ have
grown in mass due to mass transfer instead of being born this massive, or
the whole population of PCEBs is not representative of CV progenitors 
and only a small fraction of them (those with more massive WDs) may evolve into CVs.

\subsubsection{CV WD mass distribution}

\begin{figure}
   \centering
    \includegraphics[width=0.49\textwidth]{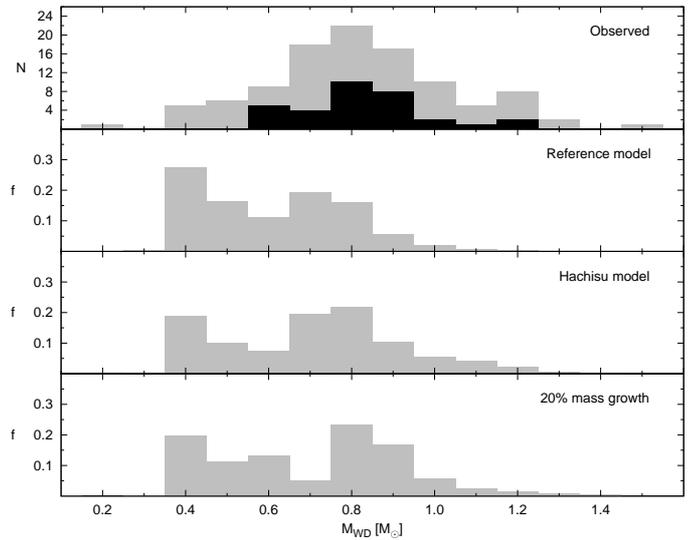}
    \caption{Mass distribution of WDs in CVs for the different models
      described in this work and the distribution derived from
      observations. \emph{From top to bottom}: observed sample used in
      \citet[][]{zorotovicetal11-1} with the black histogram representing 
a sub-sample for which the mass determination is presumably more reliable, 
distribution predicted by our reference model, distribution derived from the
reference model including a strong wind during TTMT, distribution predicted if
it is assumed that 20\% of the transferred mass during a nova cycle remains on
the WD. The fractions are normalized to the total formation probability of all
CVs in each model. 
 \label{fig:cvwd}}
\end{figure}

Figure\,\ref{fig:cvwd} shows the observed WD mass distribution and the
distributions predicted by the different calculations that we performed, which
are normalized by the sum of the formation probabilities of all CVs in each
model. We used a larger bin size than in Figs.\,\ref{fig:refmodel}
and\,\ref{fig:hachisu} to facilitate comparison with the observed
distribution. Our reference model, which is mostly based on the current
hypotheses of CV evolution, drastically disagrees with the observed
distribution (second and first panel from the top, respectively). While a small
peak at $0.8\,\Msun$ is predicted, the distribution still contains far too
many low-mass WDs and therefore also peaks at $0.4\,\Msun$, which is not
observed. The mean WD mass predicted by the reference model,
that is, $0.62\,\Msun$, clearly is too low when compared to the value derived
from observations: $\langle M_{\rm WD}\rangle = 0.83 \pm 0.02$
\citep{zorotovicetal11-1}. As shown by \citet{zorotovicetal11-1}, the high WD
masses derived from observations cannot be explained by observational biases,
and therefore it seems that an important ingredient of CV formation and/or
evolution is missing in the standard model. 

One of the possibilities proposed by \citet{zorotovicetal11-1} is to assume that a
large number of CVs descend from TTMT. Indeed, as shown in the third panel
from the top in Fig.\,\ref{fig:cvwd}, 
the mean WD mass increases to $\langle M_{\rm WD} \rangle = 0.71\,\Msun$ if a
strong wind that extends the range 
of mass-transfer rates leading to stable hydrogen burning is assumed during
TTMT. However, the distribution is still not in accordance 
with the mean mass derived from observations by \citet{zorotovicetal11-1}
i.e. $\langle M_{\rm WD}\rangle = 0.83 \pm 0.02$. 
Most importantly, there is still a
significant contribution (20.9\,\%) of low-mass WDs, in particular He WDs, to
the WD mass distribution, which is not observed
\citep{zorotovicetal11-1}. 

The second option outlined by \citet{zorotovicetal11-1} is mass growth
during the CV phase, that is, assuming that during a nova cycle the WDs gain mass
instead of slowly being eroded as predicted by the standard model. However,
while this does allow increasing the mean WD mass, none of the distributions
calculated for different fractions of mass growth is similar to the observed
one. In Fig.\,\ref{fig:cvwd} (bottom panel) we show the distribution 
assuming that 20\,\% of the transferred mass is not lost during a nova
eruption. This shows that it is in principle possible to obtain a WD mass 
distribution of CVs that is dominated by massive WDs assuming mass growth
during nova cycles, but the predicted distribution still disagrees
with the observed one. The peak at $0.8\,\Msun$ is less
dominant and less broad than in the observed distribution, and
the predicted distribution also shows a peak for He WDs at
$0.4\,\Msun$, which is not observed.  

Based on Fig.\,\ref{fig:cvwd}, none of the two options that we tested seems
to be the definitive answer for the WD mass distribution in CVs, especially regarding the
lack of systems with He WDs observed.
However, we did not perform any fine-tuning but just used standard assumptions 
plus the wind proposed by Hachisu during TTMT or allowed for mass growth
of the WD mass during the CV phase. 
As discussed in Sect.\,\ref{sec:mbce}, one could further reduce
the He WD fraction for example by reducing the CE efficiency or the critical mass
ratio for stable mass transfer in CVs. The latter option will be discussed in
a second paper. 

\subsubsection{WD masses as a function of period}
\begin{figure*}[!ht]
        \centering
        \subfloat[]{\includegraphics[width=0.3\textwidth]{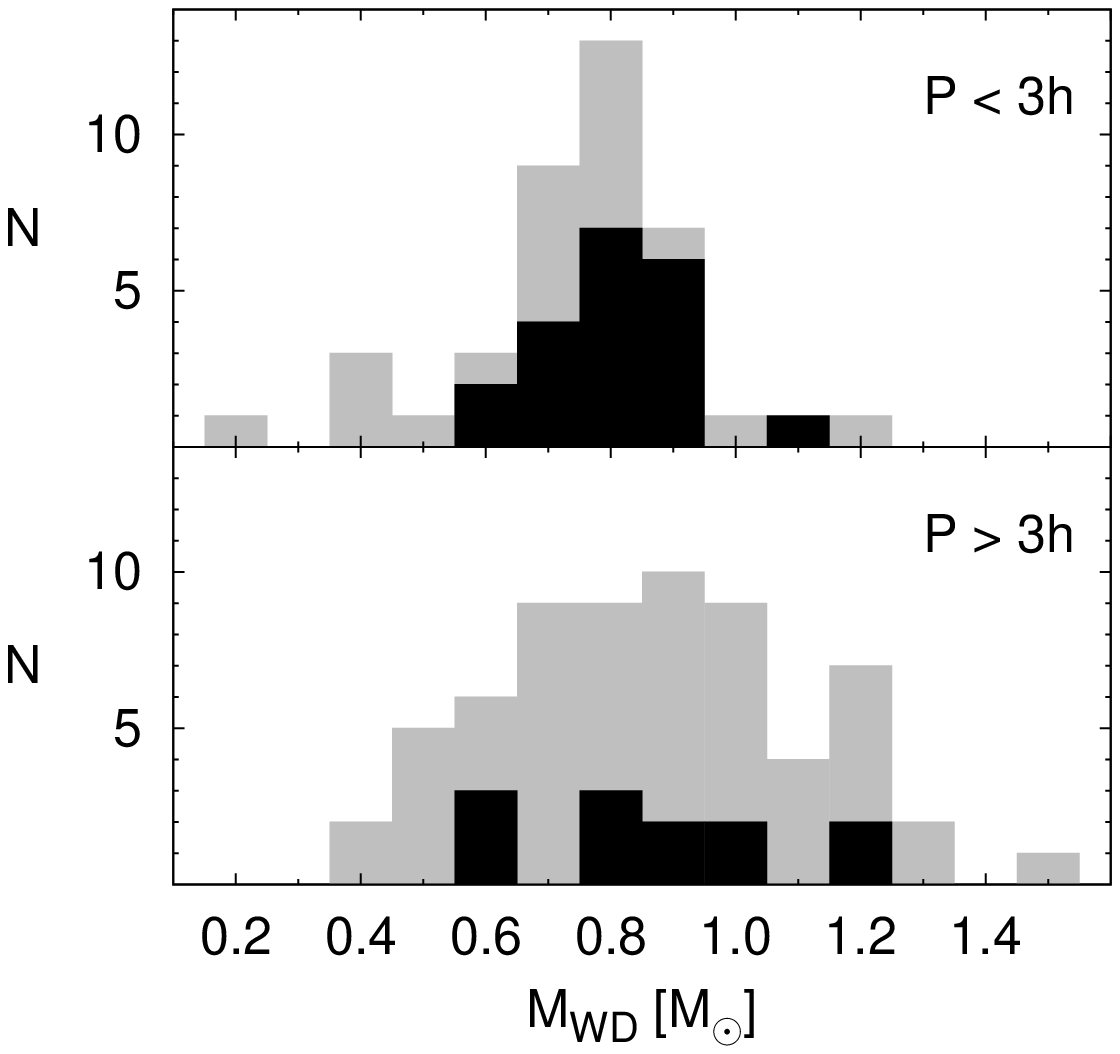}}\hfill
        \subfloat[]{\includegraphics[width=0.3\textwidth]{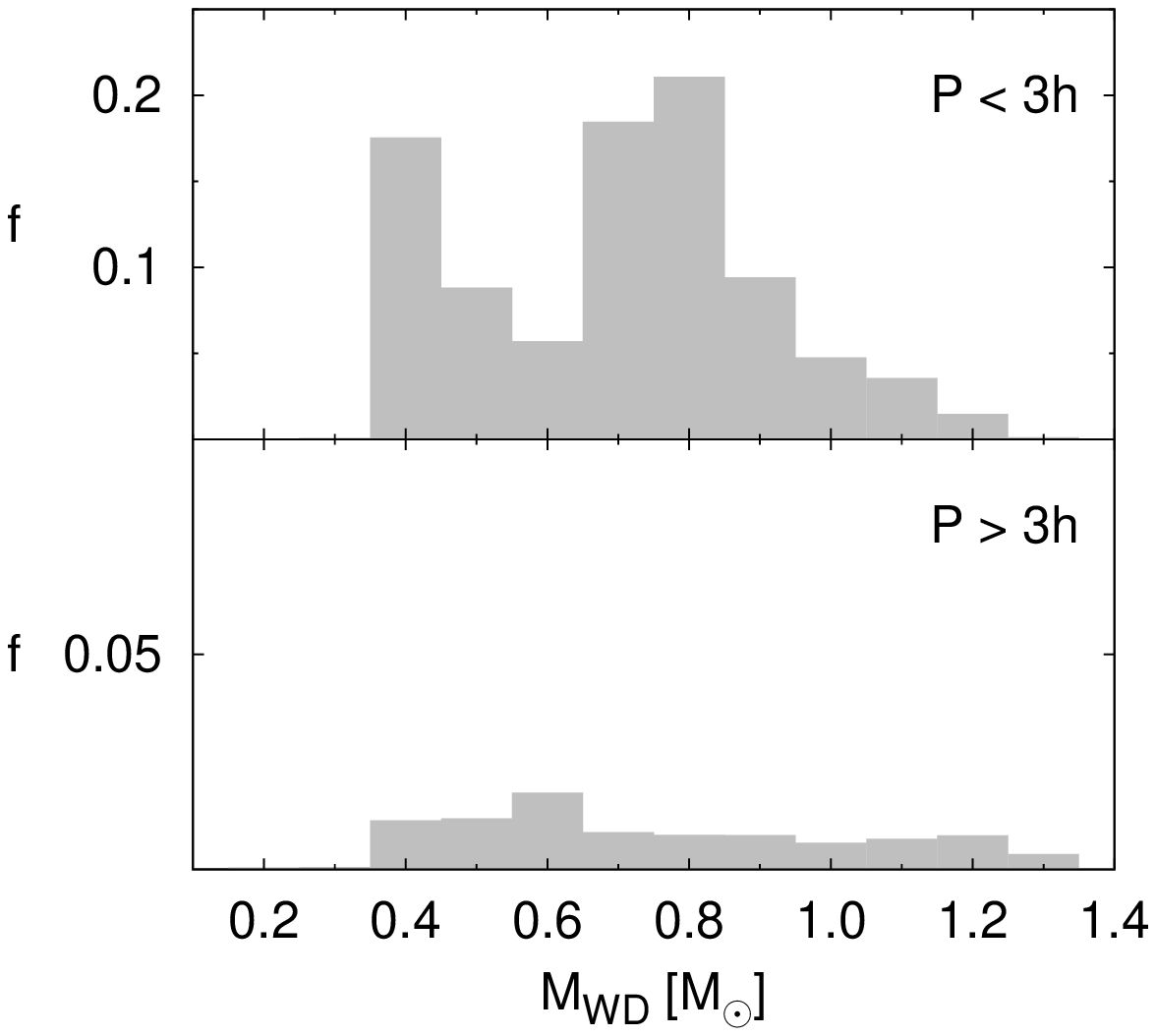}}\hfill
        \subfloat[]{\includegraphics[width=0.3\textwidth]{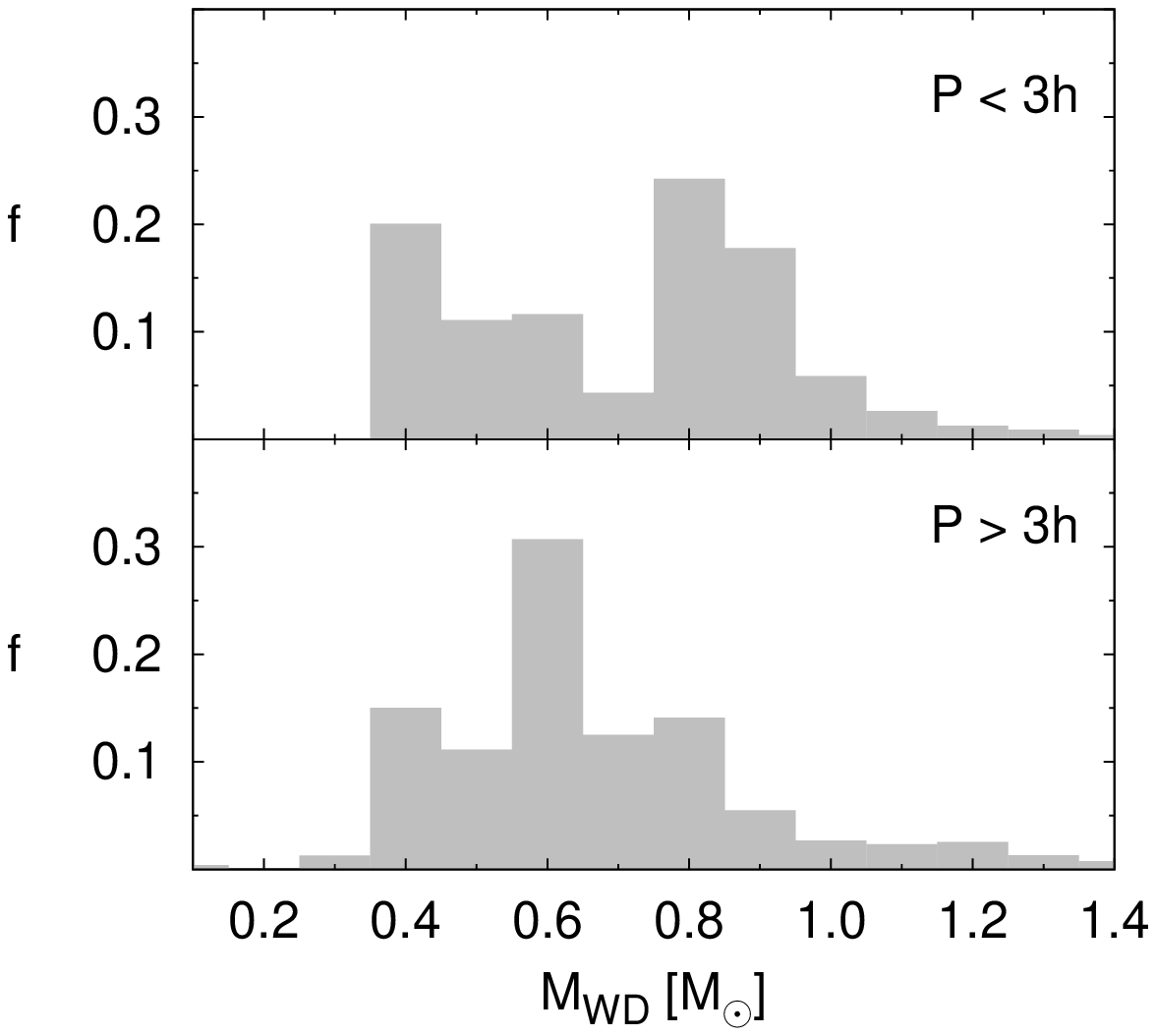}}
    \caption{WD mass distribution used by \citet{zorotovicetal11-1}
      (left) of our model with the \emph{Hachisu} wind (middle) and of our
      model including 20 per cent mass growth of the WD during a nova cycle
      (right) separated into the distribution of CVs in or below the gap
      (\emph{top panels}) and CVs above the gap (\emph{Bottom panels}). The
      black histograms in the left figure represent a sub-sample of which the
      mass determination is presumably more reliable. The fractions in the
      middle and right panel are normalized to the total formation 
      probability of all CVs in our models. The observations show a
      dispersion of WD masses above the gap, while the WD masses below the gap
      are more concentrated around $0.8\,\Msun$. Both features can also be
      seen in the model that includes the \emph{Hachisu} wind (middle), but not
      in the 20 per cent mass growth model (right).} \label{fig:gap}

\end{figure*}

\citet{zorotovicetal11-1} not only showed the general WD mass distribution of
CVs derived from observations, but also the WD mass distributions of CVs above
and below the gap. This division shows that the dispersion of WD masses above
the gap is larger than of WD masses below the gap, which strongly peaks at
$0.8\,\Msun$, while the mean masses of both sub-samples are nearly identical
(see Fig.\,\ref{fig:gap}, left).  

The best model assuming mass growth disagrees with both observed properties, 
the predicted mean mass increases towards shorter orbital periods and 
both distributions are equally peaked (see Fig.\,\ref{fig:gap}, right). 
In contrast, our model incorporating the \emph{Hachisu} wind that produces a
large fraction of CVs descending from TTMT and which best
reproduces the overall observed WD mass distribution, also shows a wide
spread of WD masses above the gap and a concentration around 0.8\,$\Msun$
below the gap (see Fig.\,\ref{fig:gap}, middle). This could indicate that a
significant fraction of post-TTMT CVs might be the missing ingredient in our
current understanding of CV evolution. However, as mentioned in the last
section, the latter model still produces far too many CVs containing He-core WDs.

\subsubsection{Fraction of CVs with evolved donors}
\citet{schenkeretal02-1} were the first to predict that the secondary stars of
CVs that went through a phase of TTMT can look different from the secondaries
of CVs that did not. First, the secondaries of post-TTMT CVs descend from
relatively massive stars and might therefore be significantly more evolved
than genuine low-mass stars. This should cause them to have later spectral
types above the gap and earlier spectral types below the gap
\citep{kolb+baraffe99-1}. The first is apparently the case for two famous long
orbital period CVs, AE Aqr and V1309 Ori. The latter has become the
standard explanation for some CVs below the period gap with very early
spectral types \citep{thorstensenetal02-1,littlefairetal06-1,thorstensen13-1}. 
Recently, such a system has been found well inside the period gap 
\citep{rebassa-mansergasetal14-1}, in agreement 
with the predictions of \citet{kolb+baraffe99-1} that CVs with evolved donors
should enter the period gap at shorter orbital periods or not at all. 

As a second possibility for identifying post-TTMT CVs, \citet{schenkeretal02-1}
predicted unusually high N/C UV line ratios for low-mass secondaries if CNO
processed material is brought to the surface by convection. Shortly after this
prediction had been made, an HST/STIS snapshot program of 31 CVs with strong
emission lines revealed that indeed four systems had extremely enhanced N/C UV
line ratios \citep{gaensickeetal03-1}.  

Thus, there is clear observational evidence for the existence of post-TTMT
CVs. However, the fraction of systems with extremely
enhanced N/C UV line flux ratios seems to be significantly smaller 
than the fraction of post-TTMT CVs suggested by 
the model including the \emph{Hachisu} wind. Thus the model represents a viable
explanation if only a fraction of post-TTMT CVs shows enhanced N/C UV
line ratios.

A second problem for the \emph{Hachisu} model is that the predicted fraction 
of CVs containing evolved donor stars
increases with the increasing number of post-TTMT CVs (see Table
\ref{tb:summary}). This number would further increase if for
instance a lower CE efficiency was
used to reduce the number of CVs with He-core WDs (which currently is too
large compared with the observations). 
In Sect.\,\ref{sec:evolved_donors} we estimated that  
the fraction of CVs with evolved donors (that evolve through the gap as
accreting systems) should not exceed $\sim20-30$ 
per cent because otherwise the period gap would be less pronounced than is
observed. 
Our simulated CV population assuming the \emph{Hachisu} wind 
predicts almost 40 per cent of CVs with evolved donors, which exceeds this
limit. We therefore conclude that assuming a large number of CVs descending 
from TTMT can at least partly explain the high WD masses in CVs 
but -- unfortunately -- generates problems with the classical explanation of the orbital period
gap.

\section{Conclusion}

We have performed binary population models of CVs with special emphasis 
on the predicted WD mass distributions with the aim to evaluate possible
solutions for the problem of the high-mass WDs in CVs identified by
\citet{zorotovicetal11-1}. We investigated the possibilities of WD mass
  growth during a preceding phase of thermal time-scale mass transfer and 
assuming net mass growth of WDs in CVs during nova cycles. 
In the latter case the predicted WD mass distributions drastically disagree with
the observed one. 
In the first case, we find the best
resemblance to the observed WD mass distribution by assuming a strong
wind that increases the range of mass transfer rates, leading to stable 
hydrogen burning (as is frequently assumed in the 
single degenerate scenario for SN\,Ia, \citealt{hachisuetal96-1}). 
This model also explains the 
observation that the dispersion of WD masses is larger above than below 
the orbital period gap. However, this model still predicts 
a large population of CVs with He WD primaries and evolved donor
stars, which clearly contradicts the observations.
We therefore conclude that WD mass growth during nova cycles or during a
preceding phase of thermal time-scale mass transfer 
cannot solve the discrepancy between the observed and predicted WD masses 
in CVs.

\begin{acknowledgements}
We are grateful to Joke Claeys, Onno Pols and Rob Izzard for valuable discussions, comments and advice.
MZ acknowledges support from CONICYT/FONDECYT/POSTDOCTORADO/3130559. MRS
thanks FONDECYT (project 1141269) and the Millennium Nucleus RC130007 (Chilean Ministry of Economy). 
\end{acknowledgements}

\bibliographystyle{aa}
\bibliography{aabib}

\end{document}